# Super-speeds with Zero-RAM:
## Next Generation Large-Scale Optimization in Your Laptop!

Mark Amo-Boateng, Ph.D[1,2]

[1]High Performance Computing Lab, Earth Observation Research and Innovation Center, University of Energy and Natural Resources, Sunyani – Ghana

[2]Enyinam Technologies Limited, P.O. Box BT 622, Community 2, Tema – Ghana

mark.amo-boateng@uenr.edu.gh, m.amoboateng@gmail.com

*Abstract*—**This article presents the novel breakthrough general purpose algorithm for very large-scale optimization problems. The novel algorithm is capable of achieving breakthrough speeds for very large-scale optimization on general purpose laptops and embedded systems. Application of the algorithm to the Griewank function was possible in up to 1 billion decision variables in double precision took only 64,485 seconds (~18 hours) to solve, while consuming 7,630 MB (7.6 GB) of RAM running on a single thread in a laptop CPU. This confirms that the algorithm is computationally and memory (space) linearly efficient, and can find the optimal or near-optimal solution in a fraction of the time and memory that many conventional algorithms require. It is envisaged that this will open up new possibilities of real-time real-world very large-scale optimization problems on personal laptops and embedded systems.**

## I. Introduction

Many problems in the advanced sciences and engineering usually involve finding the minimum (or maximum) of a certain cost function[1]–[12]. These problem formulations are usually known as optimization or calibration problems – when it involves either finding the best amongst a set of solutions or fine-tuning parameters of a system model with respect to a known reference. Recent advances have seen many real-world problems been modeled as very large-scale optimization problems which are difficult to solve by conventional optimization algorithms. Typical examples include artificial intelligence, text cluster analysis, DNA sequencing, molecular simulations, quantum chemistry, spectroscopy analysis, geophysical analysis, drug discovery, genomic research, distributed hydrological modeling, etc.

In general, the exact solutions of these current real-world optimization problems, which can usually have over $10^6$ dimensions, are believed to be hard to find because they are NP complete [12]–[16], stipulating that the computational requirements for an exact analytic solution grow exponential faster than the number of decision variables/parameters and thus cannot be solved in real (polynomial) time, even on supercomputing clusters. As such, algorithms that give a very good approximate solution in real time has been the focus of many optimization research in recent years. Even though strides have been made through meta-heuristic (nature inspired) algorithms in terms of reducing the computational requirements [12]–[14], [16], [17], the memory requirements of these algorithms make them prohibitive to run them on large scale optimization/calibration problems on conventional laptops and embedded systems. Thus, confining large-scale optimization problems only to compute accelerators, clusters, and supercomputers.

## II. Challenges with Large Scale Optimization

Despite the advances in recent meta-heuristic algorithms [12], [16], quite a number of notable challenges remain, barring efficient solution to large-scale optimization problems [18]–[20]. They include:

- *High computation complexity*: high computation complexity, usually greater than $O(N^2)$, makes the application of these algorithms limited to very small problem sizes.

- *High computational intensity*: optimization algorithms sometimes presents computational overhead far greater than the actual optimization problem, increasing the computational time required to find the optimal solution.

- *High memory complexity:* the memory requirements of many optimization algorithms is very large. This limits the application of automatic optimization methods to problems with few dimensions on typical laptops and PCs. Application to very large problems can only take place on large supercomputing clusters.

- *Curse of dimensionality:* as the number of optimization parameters increases, the parameter surfaces usually becomes ill defined due to parameter interactions. This greatly affects the ability of optimization algorithms to find the true optimum parameter set. As such, many good optimization algorithms perform poorly as the dimensionality of the optimization problem increases. This is known as the curse of dimensionality and limits the application of automatic optimization algorithms to problems with few dimensions.

- *Non-continuous and non-convex parameter surfaces:* many automatic optimization algorithms are designed and tested with benchmark functions that have convex surfaces. With the exception of algorithms that incorporate Monte Carlo methods and uncertainty analysis, other algorithms may fail to consistently find the true global optimum in problems where one or more dimensions is discontinuous or does not have a convex surface.

In the light of these challenges, it is desirable to have optimization algorithms that minimize or eliminates these challenges. Thus, the Amo-Boateng Optimization Algorithm (ABO) was developed to minimize the effect of these challenges on large-scale optimization. ABO was applied to the Griewank function up to 1 billion dimensions, and its performance metric was compared to a classical optimization algorithm, the Nelder-Mead algorithm; and a more recent state-of-the-art hybrid heuristic algorithm, the MA-SW-Chains memetic algorithm.

## III. ALGORITHMS AND BENCHMARK

The ABO algorithm, Nelder-Mead, and the MA-SW-Chains algorithms are briefly described below. This is briefly followed by a description of the Griewank Benchmark Function.

### A. Amo-Boateng Optimization Algorithm

The Amo-Boateng Optimization Algorithm (ABO) is a novel algorithm that is linear in computational and memory complexity. It is based on beliefs of how the eye visually perceives and scan's neighboring objects in fast moving situations to allow each person make the optimal decisions in real time. The general optimization problem can be defined as:

$$F = f(\{x_i\});$$
$$x_i = \{x_1, x_2, x_3, ..., x_n\}; \quad (1a)$$
$$x_i \in P_i \big|_{feasible}, P_i \big|_{feasible} \subset \mathbb{R}_i$$

Equation 1a is subject to internal dependencies (***G***) and external constraints (***H***). This is given by:

$$G_{ij} = \propto_{ij} g(x_i, u_j)\big|_{i \neq j};$$
$$u_j \in \{x_k\}\big|_{u_j \neq x_k}$$
$$H_n = h(\{u_k\}); \quad (1b)$$
$$u_k \in \{x_i\}\big|_n \quad \forall \ n \text{ constraints}$$

Where *$x_i$* is the feasible points in each parameter space *$P_i$* of the optimization problem. Thus, a generic solution is given by *{$x_i$}\** defined by:

$$F_{min} \Rightarrow f(\{x_i\}^*) \leq f(\{x_i\}) \quad (2)$$

ABO works by linear sampling the parameter spaces of the optimization problem in a manner shown in Figure 1 below. This method ensures that additional memory is not used up before computation of the objective function in each iteration, thereby making it computationally linear and memory efficient. In ABO, the only memory required for allocation is the memory for the decision variables of the optimization problem and solution store of the objective.

### B. The Nelder-Mead Algorithm

The Nelder-Mead (NM) algorithm [21]–[26] (also known as the Multi-Start Downhill Simplex method) has been extensively used for various optimization/calibration problems in science and engineering. NM has the advantage of being simple to implement and consumes very little memory. NM also forms the basis of many advanced hybrid optimization algorithms. The efficiency of NM has warranted its inclusion in the Toolkit for Advanced Optimization[27] (TAO) by Argonne National Laboratory[1] and also included in the Portable, Extensible Toolkit for Scientific Computation[2] (PETSc) for used on national supercomputing clusters such as the Titan in the USA. The Toolkit for Advanced Optimization (TAO) is aimed at the solution of large-scale optimization problems on high-performance architectures. A full description of NM, TAO and PETSc is beyond the scope of this article and can be found in the references provided. NM is used in this article for performance comparison with the ABO algorithm. NM is developed in C/C++ and maintained publicly and was obtained and used for this project

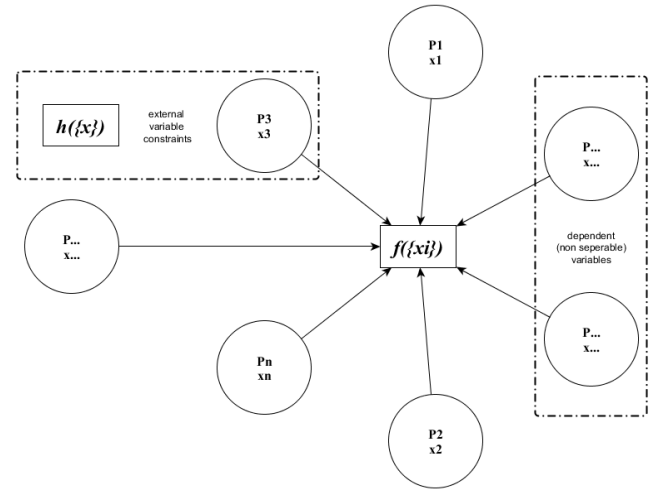

*Figure 1: The optimization problem with internal dependencies and external constraints. Arrows indicate the sampling of the parameter spaces by ABO before objective function computation in each iteration.*

### C. MA-SW-Chains Algorithm

Memetic Algorithms (MA) are hybridized Evolutionary Algorithms (EAs) and Local Search (LS) algorithms that take advantage of both exploratory and exploitation of high dimensional problems [28]. However, many LS methods do not perform well on high dimensional problems. This led to the development of an MA with LS method that performs well on both very large dimensional problems, the MA-SW-Chains algorithm [29]. MA-SW-Chains combines the classic scalable Solis Wets' algorithm with stochastic MA for continuous optimization (MA-CMA-Chains) for high dimensional problems [28], [29]. This works by chaining each global search agent to different individual local search agents based on its features.

---

[1] http://www.mcs.anl.gov/project/tao-toolkit-advanced-optimization
[2] http://www.mcs.anl.gov/petsc/

The MA-SW-Chains algorithm was adjudged the overall best and winner of the large-scale global optimization session in the IEEE Congress on Evolutionary Computation (2010) competition, outperforming known algorithms such as the DECC-CG, and MLCC algorithms [28], [29]. A parallel implementation of the MA-SW-Chains on GPUs significantly reduces the computational time required for very high dimensional problems, and their baseline results would be compared to the ABO algorithm.

*D. Griewank Benchmark Test Function*

The Griewank function is a classical optimization benchmark function for unlimited dimensions [12], [19], [30]–[33]. It has many widespread local minima, which are evenly distributed. It is usually evaluated in the domain $x_i \in [-600, 600]; i = 1 \ldots d$. It is defined by:

$$f(\boldsymbol{x}) = \sum_{i=1}^{d} \frac{x_i^2}{4000} - \prod_{i=1}^{d} \cos\left(\frac{x_i}{\sqrt{i}}\right) + 1; \quad (3)$$

With a global optimum at:
$$f(\boldsymbol{x}^*) = 0, at\ \boldsymbol{x}^* = (0, \ldots, 0) \quad (4)$$

The Griewank function for one and two dimensions are shown below (Figure 1 and 2), showing the many widespread local minima:

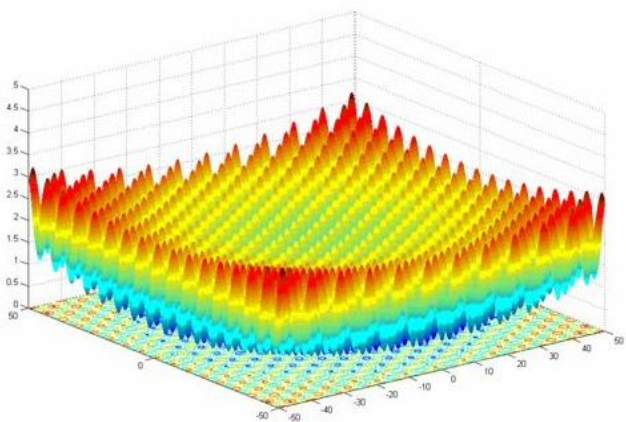

*Figure 1*: Griewank function in 2-D on the domain $x_i \in [-50, 50]$[3].

---
[3]Image source: http://www-optima.amp.i.kyoto-u.ac.jp/member/student/hedar/Hedar_files/TestGO_files/Page1905.htm

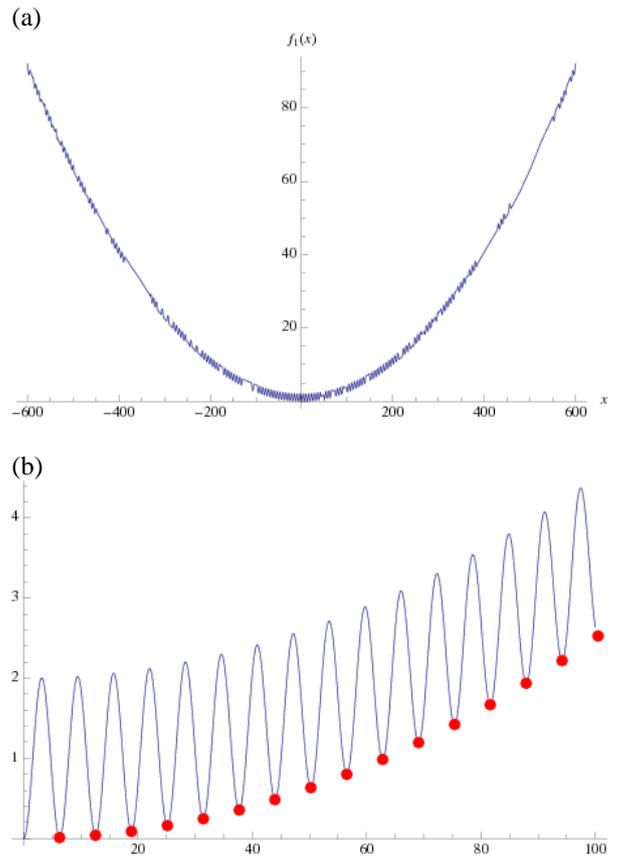

*Figure 2:* The Griewank function in 1-D. (a) on the typical domain of $x_i \in [-600, 600]$. (b) Zoomed-in of the Griewank function on the domain $x_i \in [0, 100]$[4].

IV. EXPERIMENTS WITH ABO, NM, AND MA-SW-CHAINS

ABO and NM algorithms were applied to the Griewank function for different dimensions and the following performance characteristics were recorded:

- Random Access Memory (RAM) usage
- Compute Wall Time
- Number of Function Evaluations
- Objective Function Values

The test platform for this experiment was a general-purpose laptop computer with i7-6700HQ @ 2.6 GHz with 16GB of RAM running on 128GB SSD with Windows 10 Home operating system. The development platform was Visual Studio 2015 community. The program is compiled in Release x64 bit mode to allow for more RAM usage. The experiment was compiled for single threaded applications.

*A. Random Access Memory (RAM) usage*

To assess the RAM consumed by each algorithm, Process Explorer Utility was employed to measure the amount of memory consumed by ABO or NM algorithms. The results are given in the tables 1 and 2.

---
[4] Images curtesy of Math World:
http://mathworld.wolfram.com/GriewankFunction.html

TABLE I. SINGLE PRECISION RAM USAGE BY NM AND ABO

| Dimension | NM (KB) | ABO (KB) | Theory (KB) |
|---|---|---|---|
| 2 | 436 | 436 | 0.01 |
| 10 | 432 | 438 | 0.04 |
| 100 | 480 | 436 | 0.4 |
| 1,000 | 4,368 | 432 | 4 |
| 10,000 | 391,332 | 476 | 40 |
| [a]100,000 | 5,510,120 | 820 | 400 |
| 1,000,000 | - | 4,292 | 4,000.00 |
| 10,000,000 | - | 39,500 | 40,000.00 |
| 100,000,000 | - | 391,052 | 400,000.00 |
| 1,000,000,000 | - | 3,906,688 | 4,000,000.00 |

[a] NM program crashes due to insufficient RAM; values show last recorded resource usage before program crashes.

TABLE II. DOUBLE PRECISION RAM USAGE BY NM AND ABO

| Dimension | NM (KB) | ABO (KB) | Theory (KB) |
|---|---|---|---|
| 2 | 428 | 432 | 0.02 |
| 10 | 428 | 436 | 0.08 |
| 100 | 508 | 432 | 0.80 |
| 1,000 | 8,304 | 432 | 8.00 |
| 10,000 | 782,268 | 508 | 80.00 |
| [a]100,000 | 5,828,688 | 1,208 | 800.00 |
| 1,000,000 | - | 8,248 | 8,000.00 |
| 10,000,000 | - | 78,512 | 80,000.00 |
| 100,000,000 | - | 781,640 | 800,000.00 |
| 1,000,000,000 | - | 7,965,384 | 8,000,000.00 |

[a] NM program crashes due to insufficient RAM; values show last recorded resource usage before program crashes.

The results from Table 1 and Table 2 show that ABO consumes significantly fewer memory resources than the Nelder-Mead algorithm, and it is more stable and suitable for very large-scale optimization problems.

*B. Compute Wall Time and Function Evaluations*

The length of time it takes to optimize the Griewank test function for different dimensions for ABO and NM was also assessed. The Griewank function was then optimized based on principles of the ABO algorithm and results experiments were carried out in double precision for the ABO algorithm. The optimized double precision Griewank function for ABO is represented as ABO-Opt. This experiment was carried out using a single thread and absolutely no parallelism. The associated number of function evaluations (FE), as a result of convergence or maximum iteration limit, were also recorded. The results are given in the tables 3 to 5.

TABLE III. SINGLE PRECISION WALL TIME (SECONDS)

| Dimension | NM | ABO | ABO-Opt |
|---|---|---|---|
| 2 | 0.031 | 0.004 | 0.013 |
| 10 | 4.011 | 0.005 | 0.016 |
| 100 | 29.187 | 0.121 | 0.014 |
| 1,000 | 59.886 | 4.101 | 0.031 |
| 10,000 | - | 376.782 | 0.146 |
| 100,000 | - | 33,505.231 | 1.127 |
| 1,000,000 | - | - | 10.969 |
| 10,000,000 | - | - | 108.532 |
| 100,000,000 | - | - | 1,068.721 |
| 1,000,000,000 | - | - | 64,489.001 |

TABLE IV. SINGLE PRECISION FUNCTION EVALUATIONS

| Dimension | NM | ABO | ABO-Opt |
|---|---|---|---|
| 2 | 50 | 1000 | 500 |
| 10 | 20,013 | 5,000 | 2,500 |
| 100 | 200,103 | 50,000 | 25,000 |
| 1,000 | 2,000,985 | 500,000 | 250,000 |
| 10,000 | - | 5,000,000 | 2,500,000 |
| 100,000 | - | 50,000,000 | 25,000,000 |
| 1,000,000 | - | 500 million | 250,000,000 |
| 10,000,000 | - | 5 billion | 2,500,000,000 |
| 100,000,000 | - | 50 billion | 25 billion |
| 1,000,000,000 | - | 500 billion | 250.9 billion |

TABLE V. SINGLE PRECISION BEST OBJECTIVE FUNCTION VALUES

| Dimension | NM | ABO | ABO-Opt |
|---|---|---|---|
| 2 | 11.381 | 0.00 | 3.841e-14 |
| 10 | 5.59887 | 0.071 | 1.075e-09 |
| 100 | 36.5997 | 0.009 | 5.461e-13 |
| 1,000 | 5,625.82 | 7.114e-11 | 2.644e-12 |
| 10,000 | - | 0.00021 | 8.291e-12 |
| 100,000 | - | 0.00986 | 6.081e-11 |
| 1,000,000 | - | - | 1.092e-09 |
| 10,000,000 | - | - | 5.4269e-06 |
| 100,000,000 | - | - | 1.6238e-07 |
| 1,000,000,000 | - | - | 0.0017705 |

*C. Performance ABO and MW-SW-Chains Algorithm*

The performance of the classic ABO, ABO-Opt and ABO-GPU versions of the algorithm is now compared to published results of GPU accelerated results of the MA-SW-Chains algorithm. The published MA-SW-Chains experiment performed only 500,000 function evaluations on 24 GB RAM computer with i7 CPU processor @ 2.8 GHz; whilst GPU version was on NVidia Titan GPU with G GB RAM and 2688 CUDA cores [29].

The ABO and ABO-Opt are run on single threads; ABO-GPU is run on laptop gaming GPU NVidia GTX 1060 with 6 GB RAM and 1280 CUDA cores. The results of the experiments (performed in double precision) are presented in the tables 6 and 7.

TABLE VI. 500K FUNCTION EVALUATIONS ON CPUS (SECONDS)

| Dimension | MA-SW-Chains (s) | ABO (s) | ABO-Opt (s) |
|---|---|---|---|
| 100,000 | 49,258 | 1,108.1 | 1.167 |
| 500,000 | 240,820 | 5,491.5 | 5.598 |
| 1,000,000 | 479,457 | 8,072.5 | 10.814 |
| 1,500,000 | 727,870 | 10,029.1 | 16.216 |
| 3,000,000 | 1,444,441 | 15,861.2 | 32.027 |
| 5,000,000 | - | 23,757.3 | 53.153 |
| 10,000,000 | - | 43,196.5 | 108.532 |
| 100,000,000 | - | 396,120.3 | 1,068.721 |
| 1,000,000,000 | - | 4,324,500 | 64,489.102 |

TABLE VII. 500K FUNCTION EVALUATIONS ON GPUS (SECONDS)

| Dimension | GPU MA-SW-Chains | GPU ABO | Speed-Up |
|---|---|---|---|
| 100,000 | 1,086.7 | 2.112 | 514x |
| 500,000 | 3,460.8 | 2.862 | 1,209x |
| 1,000,000 | 4,332.5 | 4.091 | 1,056x |
| 1,500,000 | 5,519.2 | 5.174 | 1,066x |
| 3,000,000 | 8,639.8 | 7.544 | 1,145x |

## V. COMPUTE AND MEMORY COMPLEXITY OF ABO

ABO algorithm was designed from ground-up to be compute and memory efficient. For single threaded applications, the best and worst for compute efficiency ($E_c$) of ABO for $N$ decision variables is $O(mN^1)$, where m is an intrinsic property dependent on the sampling rate and is defined by:

$$E_c = O(mN^1); 1 \leq m \leq k; k \propto sampling\ rate \quad (5)$$

Similarly, space (memory) efficiency ($E_m$) of ABO for single threaded applications for $N$ decision variables is:

$$E_m = O(sN^1); 1 \leq s \leq 3 \quad (6)$$

The best case occurs when the parameter spaces are uniform having the same upper and lower bounds with $s = 1$; the worst case is where each decision variable has different parameter spaces.

Theoretically, the parallel implementation of ABO reduces the compute complexity $O(mN^1)$ to $O(m)$, whilst the space complexity increases linearly by an additional $N$ from $O(sN^1)$ to $O[(s+N)N^1]$. Thus, they are given by:

$$E_{cp} = O(m^1); 1 \leq m \leq k; k = sampling\ rate \quad (7)$$
$$E_{mp} = O(sN^1); 2 \leq s \leq 5 \quad (8)$$

Thus, in general, by comparing Equation 5 and 6, the compute and space efficiency of ABO algorithm can be put in a generic form be given by:

$$E = O(\alpha N^1) \quad (9)$$

To show the efficacy of the ABO algorithm, the theoretical memory required for the Griewank function and the one used by the algorithm is measured and shown below. Also, the measured computation speed of ABO as compared to NM gives proof of its linear compute time (see Figure 4 and Figure 5).

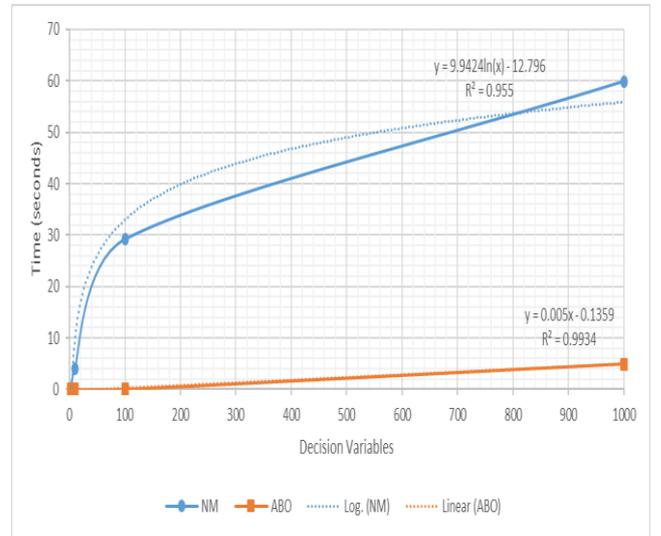

*Figure 1: Measured Computational Efficiency of NM and ABO algorithms*

The theoretical memory consumption of ABO is estimated by the bytes taken by the number of decision variables w.r.t the precision (single: 4 bytes or double: 8 bytes). Thus, for single precision: RAM = 4 bytes x decision variables. The results are shown in Table 1 and Table 2, as well as Figure 6 and Figure 7. It can further be seen that as the physical RAM limit was approached, the ABO memory usage tapered, probably because the Windows OS uses a paging system to accommodate for excess RAM demand. Also, the RAM used by ABO remained fairly constant for smaller numbers of decision variables before becoming linear at 100,000 decision variables; this is probably due to the resources used by the other components of the software.

NM algorithm memory resource usage rises quickly and crashes after 10,000 decision variables. This is because a quick analysis shows NM algorithm have a memory (space) complexity of *O[N² + 6N + 1]*, thus even single precision requires about 40 GB of RAM for 100,000 decision variables; making it impossible to run on a general laptop. Therefore, for large parameter calibration/optimization, NM is on plausible on supercomputing clusters with high RAM availability. This contrast sharply with ABO which needs only 400 KB of RAM for 100,000 parameters.

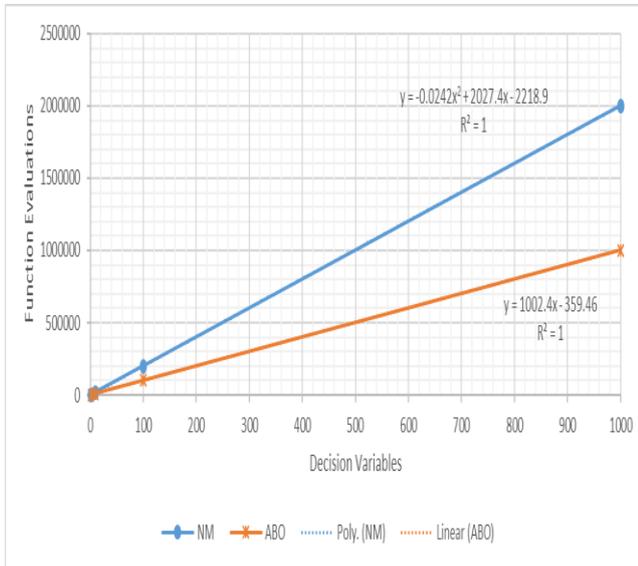

*Figure 2: Number of Function Evaluations to Convergence of NM and ABO algorithms*

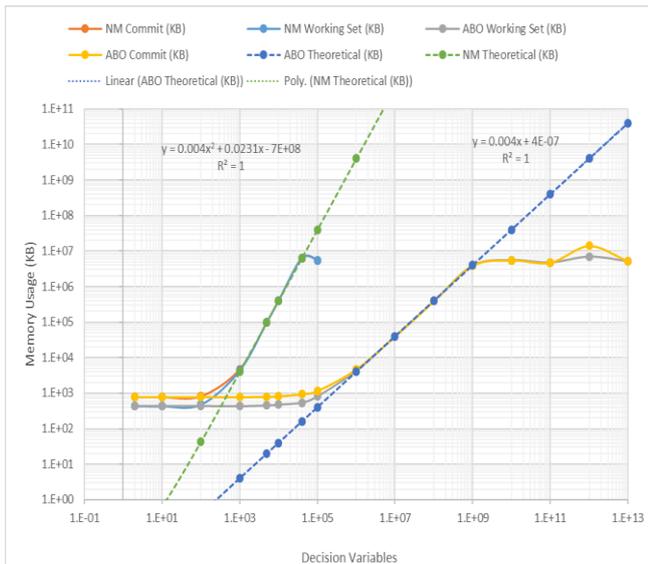

*Figure 3: Measured Single Precision Memory Resource Usage of NM and ABO on Griewank Function*

The results from above go to prove that the ABO algorithm is both computationally and memory linearly efficient. It is hoped that the algorithm can be applied to many optimization functions and applications.

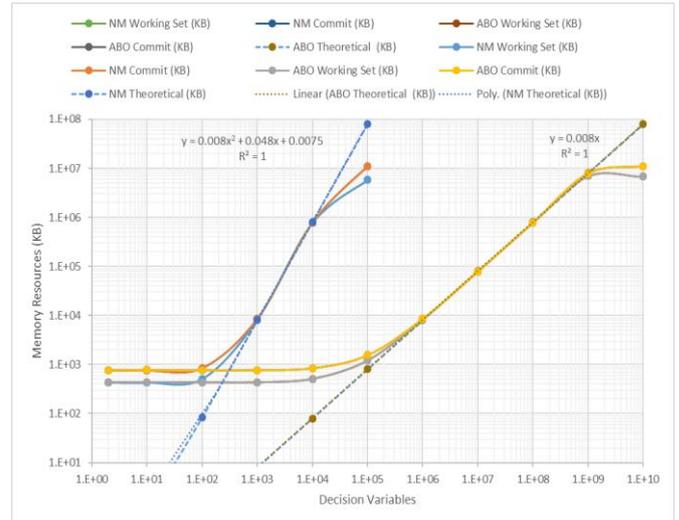

*Figure 4: Measured Double Precision Memory Resource Usage of NM and ABO on Griewank Function*

## VI. CONCLUSION

Amo-Boateng Optimization Algorithm (ABO) is a novel compute and memory efficient algorithm for optimization/calibration that was developed to address the inherent challenges faced by current algorithms in large and very large scale problems. In particular, given that recent advances in computing hardware and our general understanding of our environment have led to the development of very large-scale optimization/calibration models, the need for very fast efficient algorithms is imperative.

Amongst other challenges stated in this article, these models, however, can only be solved on HPC clusters and supercomputers, given that the current existing optimization algorithms require very large memory resources, and are computationally intensive. A novel algorithm (ABO) that is both compute and space linearly efficient, *O(αN¹)*, has been developed. This paper shows how the Wall Wait Time and Memory Resources used proves that ABO is efficient linearly. In particular, by comparing the memory used by ABO to the theoretical minimum memory resources that can be used, and finding these to be similar. The linear efficiency of ABO opens up compute speeds that are only available in supercomputing clusters and allows for solving larger problem sizes on ordinary laptops and embedded systems.

It is the author's hope that ABO will be useful in all fields of science, engineering, drug research, finance, artificial intelligence, etc. and will accelerate the time to discovery, prototyping, and market of new developments in these fields. It is deemed that with the widespread adoption of ABO, ***Super speeds with Zero-RAM*** can be achieved lowering barriers of entry and accelerating the pace of innovation in various fields. This is because the novel algorithm allows those speeds to be attained with zero additional RAM, except those required for storing the solution.

REFERENCES

[1] D. W. Hillis, "Optimization problems," Nature, vol. 330, no. 5, pp. 27–28, 1987.
[2] A. Delévacq, P. Delisle, M. Gravel, and M. Krajecki, "Parallel Ant Colony Optimization on Graphics Processing Units," J. Parallel Distrib. Comput., vol. 73, no. 1, pp. 52–61, Jan. 2013.
[3] J. Parker, U. Kim, P. Kitanidis, M. Cardiff, X. Liu, and G. Beyke, "Stochastic cost optimization of DNAPL remediation – Method description and sensitivity study," Environ. Model. Softw., vol. 38, pp. 74–88, Dec. 2012.
[4] J. Yan, H. Tiesong, H. Chongchao, W. Xianing, and G. Faling, "A shuffled complex evolution of particle swarm optimization algorithm," in Adaptive and Natural Computing Algorithms, 2007, pp. 341–349.
[5] Y. Tang, P. Reed, and T. Wagener, "How effective and efficient are multiobjective evolutionary algorithms at hydrologic model calibration?," Hydrol. Earth Syst. Sci., vol. 10, no. 2, pp. 289–307, 2006.
[6] G. F. Laniak, G. Olchin, J. Goodall, A. Voinov, M. Hill, P. Glynn, G. Whelan, G. Geller, N. Quinn, M. Blind, S. Peckham, S. Reaney, N. Gaber, R. Kennedy, and A. Hughes, "Integrated environmental modeling: A vision and roadmap for the future," Environ. Model. Softw., vol. 39, pp. 3–23, Jan. 2013.
[7] W. Wenzel and K. Hamacher, "Adaptation in Stochastic tunneling global optimization of complex potential energy landscapes," EPL (Europhysics Lett., vol. 74, no. 6, p. 944, Apr. 2006.
[8] C. Schulz, "Efficient local search on the GPU—Investigations on the vehicle routing problem," J. Parallel Distrib. Comput., vol. 73, no. 1, pp. 14–31, Jan. 2013.
[9] B. A. Berg, "Locating global minima in optimization problems by a random-cost approach," Nature, vol. 361, pp. 708–710, 1993.
[10] M. Mahdavi, M. Fesanghary, and E. Damangir, "An improved harmony search algorithm for solving optimization problems," Appl. Math. Comput., vol. 188, no. 2, pp. 1567–1579, May 2007.
[11] K. S. Lee and Z. W. Geem, "A new structural optimization method based on the harmony search algorithm," Comput. Struct., vol. 82, no. 9–10, pp. 781–798, Apr. 2004.
[12] W. Chu, X. Gao, and S. Sorooshian, "A new evolutionary search strategy for global optimization of high-dimensional problems," Inf. Sci. (Ny)., vol. 181, no. 22, pp. 4909–4927, 2011.
[13] Z. Yang, K. Tang, and X. Yao, "Large scale evolutionary optimization using cooperative coevolution," Inf. Sci. (Ny)., vol. 178, no. 15, pp. 2985–2999, Aug. 2008.
[14] Y. Hung and W. Wang, "Accelerating parallel particle swarm optimization via GPU," Optim. Methods Softw., vol. 27, no. 1, pp. 33–51, Feb. 2012.
[15] D. Lu, M. Ye, M. C. Hill, E. P. Poeter, and G. P. Curtis, "A computer program for uncertainty analysis integrating regression and Bayesian methods," Environ. Model. Softw., vol. 60, pp. 45–56, Oct. 2014.
[16] S. Navlakha and Z. Bar-joseph, "Algorithms in nature : the convergence of systems biology and computational thinking," Mol. Syst. Biol., vol. 7, no. 546, pp. 1–11, 2011.
[17] F. Kang, J. Li, and H. Li, "Artificial bee colony algorithm and pattern search hybridized for global optimization," Appl. Soft Comput., vol. 13, no. 4, pp. 1781–1791, Apr. 2013.
[18] J. a. Vrugt, H. V. Gupta, W. Bouten, and S. Sorooshian, "A Shuffled Complex Evolution Metropolis algorithm for optimization and uncertainty assessment of hydrologic model parameters," Water Resour. Res., vol. 39, no. 8, p. n/a-n/a, Aug. 2003.
[19] Q. Y. Duan, V. K. Gupta, and S. Sorooshian, "Shuffled complex evolution approach for effective and efficient global minimization," J. Optim. Theory Appl., vol. 76, no. 3, pp. 501–521, Mar. 1993.
[20] N. Muttil, S. Y. Liong, and O. Nesterov, "A Parallel Shuffled Complex Evolution Model Calibrating Algorithm to Reduce Computational Time," in MODSIM 2007: International Congress On Modelling And Simulation: Land, Water, And Environmental Management: Integrated Systems For Sustainability (2007), 2007, pp. 1940–1946.
[21] N. Pham and B. M. Wilamowski, "Improved Nelder Mead ' s Simplex Method and Applications," vol. 3, no. 3, pp. 55–63, 2011.
[22] S. Singer, "Complexity Analysis of Nelder – Mead Search," pp. 185–196, 1999.
[23] K. I. M. McKinnon, "Convergence of the Nelder--Mead Simplex Method to a Nonstationary Point," SIAM J. Optim., vol. 9, no. 1, pp. 148–158, Jan. 1998.
[24] J. A. Nelder and R. Mead, "A simplex method for function minimization," Comput. J., vol. 7, no. 4, pp. 303–318, Jan. 1965.
[25] T. Ye and S. Kalyanaraman, "A Recursive Random Search Algorithm for Black-box Optimization," no. x, pp. 1–24.
[26] D. Xu, W. Wang, K. Chau, C. Cheng, and S. Chen, "Comparison of three global optimization algorithms for calibration of the Xinanjiang model parameters," J. Hydroinformatics, 2012.
[27] T. Munson, J. Sarich, S. Wild, S. Benson, and L. C. McInnes, "TAO 2.0 Users Manual," ANL/MCS-TM-322, 2012.
[28] D. Molina, M. Lozano and F. Herrera, "MA-SW-Chains: Memetic algorithm based on local search chains for large scale continuous global optimization," IEEE Congress on Evolutionary Computation, Barcelona, 2010, pp. 1-8. doi: 10.1109/CEC.2010.5586034
[29] Miguel Lastra, Daniel Molina, José M. Benítez, "A high performance memetic algorithm for extremely high-dimensional problems", Information Sciences, Volume 293, 2015, Pages 35-58, ISSN 0020-0255, http://dx.doi.org/10.1016/j.ins.2014.09.018.
[30] M. Zambrano-Bigiarini and R. Rojas, "A model-independent Particle Swarm Optimisation software for model calibration," Environ. Model. Softw., vol. 43, pp. 5–25, May 2013.
[31] F. Zhao and J. Zhang, "An Improved Shuffled Complex Evolution Algorithm and Its Performance Analysis," Journal Comput. Inf. Syst., vol. 20, no. 61064011, pp. 8495–8502, 2012.
[32] X. Li, K. Tang, M. N. Omidvar, Z. Yang, and K. Qin, "Benchmark Functions for the CEC ' 2013 Special Session and Competition on Large-Scale Global Optimization," Australia, 2013.
[33] B. a. Tolson and C. a. Shoemaker, "Dynamically dimensioned search algorithm for computationally efficient watershed model calibration," Water Resour. Res., vol. 43, no. 1, p. n/a-n/a, Jan. 2007.